\documentclass[onecolumn]{aastex63}
\usepackage{placeins}
\usepackage{amsmath}

\received{..., 2021}
\revised{..., 2021}

\accepted{in ApJ}

\shorttitle{SiO jets in Class\,I protostars}
\shortauthors{Dutta et al.}

\begin{document}

\title{Detection of a dense SiO jet in the evolved protostellar phase}

\correspondingauthor{Somnath Dutta}
\email{sdutta@asiaa.sinica.edu.tw, cflee@asiaa.sinica.edu.tw}

\author[0000-0002-2338-4583]{Somnath Dutta}
\affiliation{Institute of Astronomy and Astrophysics, Academia Sinica, Roosevelt Rd, Taipei 10617, Taiwan, R.O.C.}

\author[0000-0002-3024-5864]{Chin-Fei Lee}
\affiliation{Institute of Astronomy and Astrophysics, Academia Sinica, Roosevelt Rd, Taipei 10617, Taiwan, R.O.C.}

\author[0000-0002-6773-459X]{Doug Johnstone}
\affiliation{National Research Council of Canada, Herzberg, Astronomy and Astrophysics Research Centre, 5071 West Saanich Road, V9E 2E7 Victoria (BC), Canada}
\affiliation{Department of Physics and Astronomy, University of Victoria, Victoria, BC V8P 5C2, Canada}

\author[0000-0002-5286-2564]{Tie Liu}
\affiliation{Shanghai Astronomical Observatory, Chinese Academy of Sciences, 80 Nandan Road, Shanghai 200030, China}

\author[0000-0001-9304-7884]{Naomi Hirano}
\affiliation{Institute of Astronomy and Astrophysics, Academia Sinica, Roosevelt Rd, Taipei 10617, Taiwan, R.O.C.}

\author[0000-0003-4603-7119]{Sheng-Yuan Liu}
\affiliation{Institute of Astronomy and Astrophysics, Academia Sinica, Roosevelt Rd, Taipei 10617, Taiwan, R.O.C.}

\author{Jeong-Eun Lee} 
\affil{School of Space Research, Kyung Hee University, Yongin-Si, Gyeonggi-Do 17104, Republic of Korea}

\author[0000-0001-8385-9838]{Hsien Shang}
\affiliation{Institute of Astronomy and Astrophysics, Academia Sinica, Roosevelt Rd, Taipei 10617, Taiwan, R.O.C.}

\author[0000-0002-8149-8546]{Ken'ichi Tatematsu}
\affil{Nobeyama Radio Observatory, National Astronomical Observatory of Japan, 
National Institutes of Natural Sciences, 
462-2 Nobeyama, Minamimaki, Minamisaku, Nagano 384-1305, Japan}
\affiliation{Department of Astronomical Science,
SOKENDAI (The Graduate University for Advanced Studies),
2-21-1 Osawa, Mitaka, Tokyo 181-8588, Japan}

\author[0000-0003-2412-7092]{Kee-Tae Kim}
\affil{Korea Astronomy and Space Science Institute (KASI), 776 Daedeokdae-ro, Yuseong-gu, Daejeon 34055, Republic of Korea}
\affil{University of Science and Technology, Korea (UST), 217 Gajeong-ro, Yuseong-gu, Daejeon 34113, Republic of Korea}

\author[0000-0002-4393-3463]{Dipen Sahu}
\affiliation{Institute of Astronomy and Astrophysics, Academia Sinica, Roosevelt Rd, Taipei 10617, Taiwan, R.O.C.}

\author[0000-0002-7125-7685]{Patricio Sanhueza} 
\affiliation{National Astronomical Observatory of Japan, National Institutes of Natural Sciences, 2-21-1 Osawa, Mitaka, Tokyo 181-8588, Japan}
\affiliation{Department of Astronomical Science,
SOKENDAI (The Graduate University for Advanced Studies),
2-21-1 Osawa, Mitaka, Tokyo 181-8588, Japan}

\author{James Di Francesco}
\affiliation{National Research Council of Canada, Herzberg, Astronomy and Astrophysics Research Centre, 5071 West Saanich Road, V9E 2E7 Victoria (BC), Canada}

\author{Kai-Syun Jhan}
\affiliation{Institute of Astronomy and Astrophysics, Academia Sinica, Roosevelt Rd, Taipei 10617, Taiwan, R.O.C.}

\author{Chang Won Lee}
\affil{Korea Astronomy and Space Science Institute (KASI), 776 Daedeokdae-ro, Yuseong-gu, Daejeon 34055, Republic of Korea}
\affil{University of Science and Technology, Korea (UST), 217 Gajeong-ro, Yuseong-gu, Daejeon 34113, Republic of Korea}

\author{Woojin Kwon}
\affil{Department of Earth Science Education, Seoul National University, 1 Gwanak-ro, Gwanak-gu, Seoul 08826, Republic of Korea}
\affil{SNU Astronomy Research Center, Seoul National University, 1 Gwanak-ro, Gwanak-gu, Seoul 08826, Republic of Korea}

\author[0000-0003-1275-5251]{Shanghuo Li}
\affiliation{Korea Astronomy and Space Science Institute (KASI), 776 Daedeokdae-ro, Yuseong-gu, Daejeon 34055, Republic of Korea}

\author[0000-0002-9574-8454]{Leonardo Bronfman}
\affil{Departamento de Astronomía, Universidad de Chile, Casilla 36-D, Santiago, Chile}

\author[0000-0003-3343-9645]{Hong-li Liu}
\affil{Department of Astronomy, Yunnan University, Kunming 650091, People’s Republic of China}

\author{Alessio Traficante}
\affil{IAPS-INAF, via Fosso del Cavaliere 100, I-00133, Rome, Italy}

\author[0000-0002-4336-0730]{Yi-Jehng Kuan}
\affiliation{Department of Earth Sciences, National Taiwan Normal University, Taipei, Taiwan, R.O.C.}
\affiliation{Institute of Astronomy and Astrophysics, Academia Sinica, Roosevelt Rd, Taipei 10617, Taiwan, R.O.C.}

\author{Shih-Ying Hsu}
\affiliation{Institute of Astronomy and Astrophysics, Academia Sinica, Roosevelt Rd, Taipei 10617, Taiwan, R.O.C.}

\author{Anthony Moraghan}
\affiliation{Institute of Astronomy and Astrophysics, Academia Sinica, Roosevelt Rd, Taipei 10617, Taiwan, R.O.C.}

\author{Chun-Fan Liu}
\affiliation{Institute of Astronomy and Astrophysics, Academia Sinica, Roosevelt Rd, Taipei 10617, Taiwan, R.O.C.}

\author{David Eden}
\affiliation{Astrophysics Research Institute, Liverpool John Moores University, IC2, Liverpool Science Park, 146 Brownlow Hill, Liverpool, L3 5RF, UK}


\author[0000-0002-6386-2906]{Archana Soam}
\affil{SOFIA Science Center, Universities Space Research Association, NASA Ames Research Center, Moffett Field, California 94035, USA}

\author{Qiuyi Luo}
\affiliation{Shanghai Astronomical Observatory, Chinese Academy of Sciences, 80 Nandan Road, Shanghai 200030, China}

\author{ALMASOP Team}

\begin{abstract} 
Jets and outflows trace the accretion history of protostars. High-velocity molecular jets have been observed from several protostars in the early Class\,0 phase of star formation,  detected with the high-density tracer SiO. Until now, no clear jet has been detected with SiO emission from isolated evolved Class\,I protostellar systems. We report a prominent dense SiO jet from a Class\,I source G205S3 (HOPS\,315: T$_{bol}$ $\sim$ 180 K, spectral index $\sim$ 0.417), with a moderately high  mass-loss rate ($\sim$ 0.59 $\times$ 10$^{-6}$ M$_\sun$ yr$^{-1}$) estimated from CO emission. Together, these features suggest that G205S3 is still in a high accretion phase, similar to that expected of Class\,0 objects.
We compare G205S3 to a representative Class\,0 system G206W2 (HOPS\,399) and literature Class\,0/I sources to explore the possible explanations behind the SiO emission seen at the later phase. We estimate a high inclination angle ($\sim$ 40$\degr$) for G205S3 from CO emission, which may expose the infrared emission from the central core and mislead the spectral classification. However, the compact 1.3\,mm continuum, C$^{18}$O emission, location in the bolometric luminosity to sub-millimeter fluxes diagram, outflow force ($\sim$ 3.26 $\times$ 10$^{-5}$ M$_\sun$km s$^{-1}$/yr) are also analogous to that of Class\,I systems. We thus consider G205S3 to be at the very early phase of Class\,I, and in the late phase of ``high-accretion". The episodic ejection could be due to the presence of an unknown binary, a planetary companion, or dense clumps, where the required mass for such high accretion could be supplied by a massive circumbinary disk.

\end{abstract}

\keywords{Low mass stars (2050); Stellar jets (1607); Stellar winds (1636); Protostars (1302); Astrochemistry (75); Star formation (1569); Stellar mass loss (1613); Young stellar objects (1834); Stellar evolution (1599)} 

\section{Introduction} \label{sec:intro}
During the earliest phases of star-formation, jets and outflows play a crucial role in mediating protostellar accretion. Previous observations suggest that jets can efficiently remove the excess angular momentum from the surfaces of circumstellar disks and allow material to fall onto the central sources \citep[see reviews by][and references therein]{2016ARA&A..54..491B,2020A&ARv..28....1L}. Therefore, jets may delineate the accretion history of protostars. Despite this important role, the jet launching timescale over protostellar evolution is not yet well constraint.

The protostellar spectral classes, 0 and I,  are observationally defined and not clearly distinct in terms of evolution. Protostars with bolometric temperature T$_{bol}$ $>$ 70 K are classified as Class I \citep{1995ApJ...445..377C}, and typically have spectral indexes, $\alpha_{IR}$ $>$ 0.3 \citep[][]{2016ApJS..224....5F}. The Class\,0 sources are embedded within dense envelopes and have sufficient surrounding material to exhibit typically very high accretion rates onto the protostar. During this phase,  protostars usually show outflows with very high mass-loss rates $\sim$ $10^{-6}$ - $10^{-7}$ M$_\sun$ yr$^{-1}$, which could produce high density jets (5-10 $\times$ 10$^{6}$ cm$^{-3}$) \citep[][]{2013A&A...551A...5E,2020A&ARv..28....1L}. Due to its high critical density, the SiO (5-4) molecular transition is the  most commonly observed tracer of such high density material \citep[][]{2004ApJ...603..198G,2015A&A...581A..85P,2021A&A...648A..45P}.  With time, the  envelope material reduces and the protostar moves from Class\,0 through Class\,I, and on to the Class\,II phase. Both the accretion and mass-loss rates typically decrease with protostellar evolution.  Therefore, SiO emission in jets is usually seen in the Class\,0 sources. Faint SiO emission is also detected along the jet axis of a few transition Class\,0-to-Class\,I sources \citep[][]{2021A&A...648A..45P}.  A Class\,I protostellar system, SVS13A was found with SiO knots in previous observations \citep[][]{2000A&A...362L..33B,2017A&A...604L...1L}. However,  SVS13A is located in a multiple system HH\,7/11 with a few outflow/jets in the region, and is a binary protostellar system composed of VLA4A and VLA4B, where VLA4B is identified as the base of the jet \citep[][]{2017A&A...604L...1L}. The spectral classification of the components of such a binary/multiple based on infrared observations could be largely affected by the multiplicity. 
No clear jets with SiO emission have been previously observed in an isolated source in the evolved Class\,I phase.

The gas content of the jets transitions from being  predominantly molecular to mostly atomic, during evolution from Class\,0 to Class\,II. 
The jets in the younger sources, such as those in the Class\,0 phase, are predominantly detected via molecular gas tracers, e.g., CO, SiO, and SO in the  (sub)millimeter and H$_2$ at infrared wavelengths. Conversely, in the older Class\,I, and II populations,  the jets are mainly traced by the atomic and ionized gas, e.g., O, H$\alpha$, and S\,II \citep{2016ARA&A..54..491B,2020A&ARv..28....1L}.

In this paper, we report the surprising detection of a Class\,I source with a clear protostellar jet seen in SiO by the Atacama Large Millimeter/submillimeter Array (ALMA). The system G205.46-14.56S3 (hereafter, G205S3) has T$_{bol}$ = 180$\pm$33 K and L$_{bol}$ = 6.4$\pm$2.4 L$_\sun$ \citep[HOPS\,315;][]{2016ApJS..224....5F,2020ApJS..251...20D}. 
We compare this system with ALMA observations of a representative Class\,0 protostar, G206.93-16.61W2 (hereafter, G206W2), with T$_{bol}$ = 31$\pm$10 K and L$_{bol}$ = 6.3$\pm$3.0 L$_\sun$  \citep[HOPS\,399:][]{2016ApJS..224....5F,2020ApJS..251...20D}. Both sources exhibit very high mass-loss rates; however, the SiO jets present very different inclination angles.  We first quantify the outflow characteristics and then discuss the evolutionary state of these two systems in terms of their observed sky orientation.

\section{Observations} \label{sec:dataObservations}

The ALMA observations of G205S2 and G206W2 were performed as part of the ALMASOP project (Project ID:2018.1.00302.S; PI: Tie Liu) in Band\,6 during Cycle\,6, from 2018 October to 2019 January, toward 72 fields  \citep[see][for more details on the ALMASOP]{2020ApJS..251...20D}.
This paper utilizes the low/high-velocity outflow tracer CO J=2-1 (230.53800 GHz), the high-velocity jet tracer SiO J=5-4 (217.10498 GHz), the envelope tracer C$^{18}$O J=2-1 (219.56035 GHz), and 1.3\,mm dust continuum emission. The acquired visibility data were calibrated using the standard pipeline in CASA 5.4 \citep{2007ASPC..376..127M}. The CO J=2-1, SiO J=5-4, and C$^{18}$O J=2-1 emission maps were created with the TCLEAN task using  a robust weighting factor of $+$0.5 on a combination of three visibility datasets (i.e., TM1+TM2+ACA) providing typical synthesized beam sizes of $\sim$ 0$\farcs$37 $\times$ 0$\farcs$32,  $\sim$ 0$\farcs$41 $\times$ 0$\farcs$34 and $\sim$ 0$\farcs$39 $\times$ 0$\farcs$32, respectively. The velocity resolution is 1.4 km\,s$^{-1}$. 
The continuum maps were created using TCLEAN and a $+$0.5 robust weighting down to a threshold of 3\,$\sigma$ theoretical sensitivity with the synthesized beam size of $\sim$ 0$\farcs$38 $\times$ 0$\farcs$34. More details on these observations and analyses are presented in \citet[][]{2020ApJS..251...20D}.

\begin{figure*}
\fig{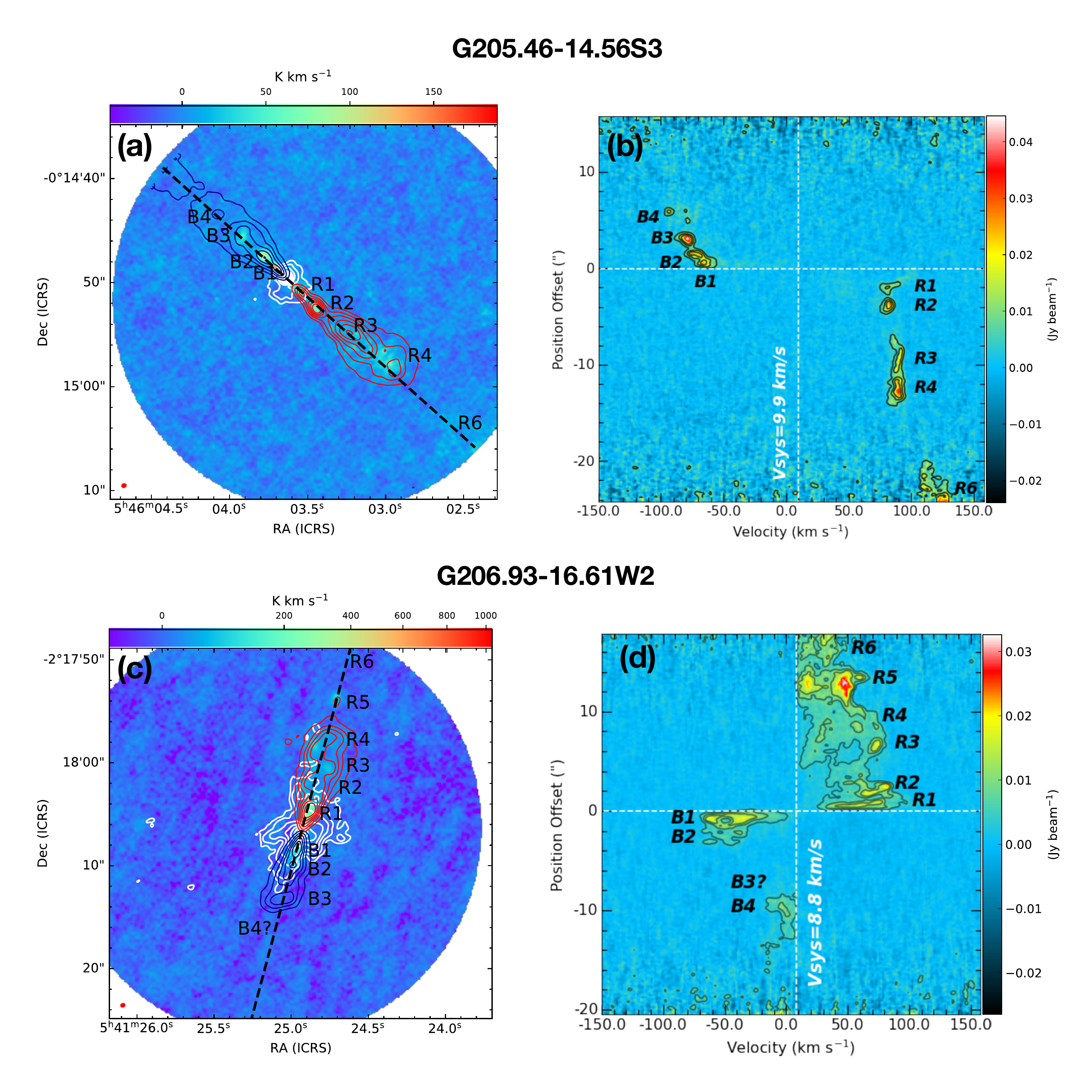}{0.9\textwidth}{}
\caption{
(a) ALMA SiO J=5-4 maps of G205S3 jet at spatial resolution of 140 AU with sensitivity of $\sim$ 9.5 K. High-velocity $^{12}$CO(2-1) contours are overplotted for blueshifted (blue) and redshifted (red) lobes at 3$\times$(1, 2, 3, 6, 9)$\sigma$, where $\sigma$ = 51 mJy\,beam$^{-1}$. The continuum contours (white) are at $\sum_{{n=0, 1, 2, 3, ..}}$3$\times$2$^n$ $\sigma$, where $\sigma$ = 80 $\mu$Jy\,beam$^{-1}$. The beam size of $\sim$ 0$\farcs$41 $\times$ 0$\farcs$34 for SiO emission is shown in red on lower left. The black dashed line indicates the jet axis.  
(b) Position-Velocity (PV) diagrams across the jet axis. The vertical dashed lines indicate the systemic velocity. The contour levels start from 3$\sigma$ with steps of 3$\sigma$, where $\sigma$ = 2 mJy\,beam$^{-1}$. The location of knots is marked in the blue lobe (B1, B2, ...) and red lobe (R1, R2, ...). 
(c) ALMA SiO J=5-4 maps of G2056W2 with sensitivity of $\sim$ 12 K. All the symbols are the same as panel `a'. The higher velocity $^{12}$CO(2-1) have sensitivity $\sigma$ = 53 mJy\,beam$^{-1}$.  The continuum emission have sensitivity $\sigma$ = 150 $\mu$Jy\,beam$^{-1}$.   
(d) PV diagram across the jet axis. All the symbols are the same as panel `b'. The sensitivity of the map is  $\sigma$ = 1.3 mJy\,beam$^{-1}$.
}
\label{fig:G205S3_G206W2SiO_jet}
\end{figure*}

\section{RESULTS}\label{sec:resultsOutflowJet}
\subsection{Detection of molecular jet in SiO emission}\label{sec:results_SiO}
Figures \ref{fig:G205S3_G206W2SiO_jet}a and c  display ALMA maps of the two jets in SiO (5-4) at $\sim$ 140 au resolution along with the high-velocity CO outflow contours over plotted in blue and red. 
\begin{figure*}
\fig{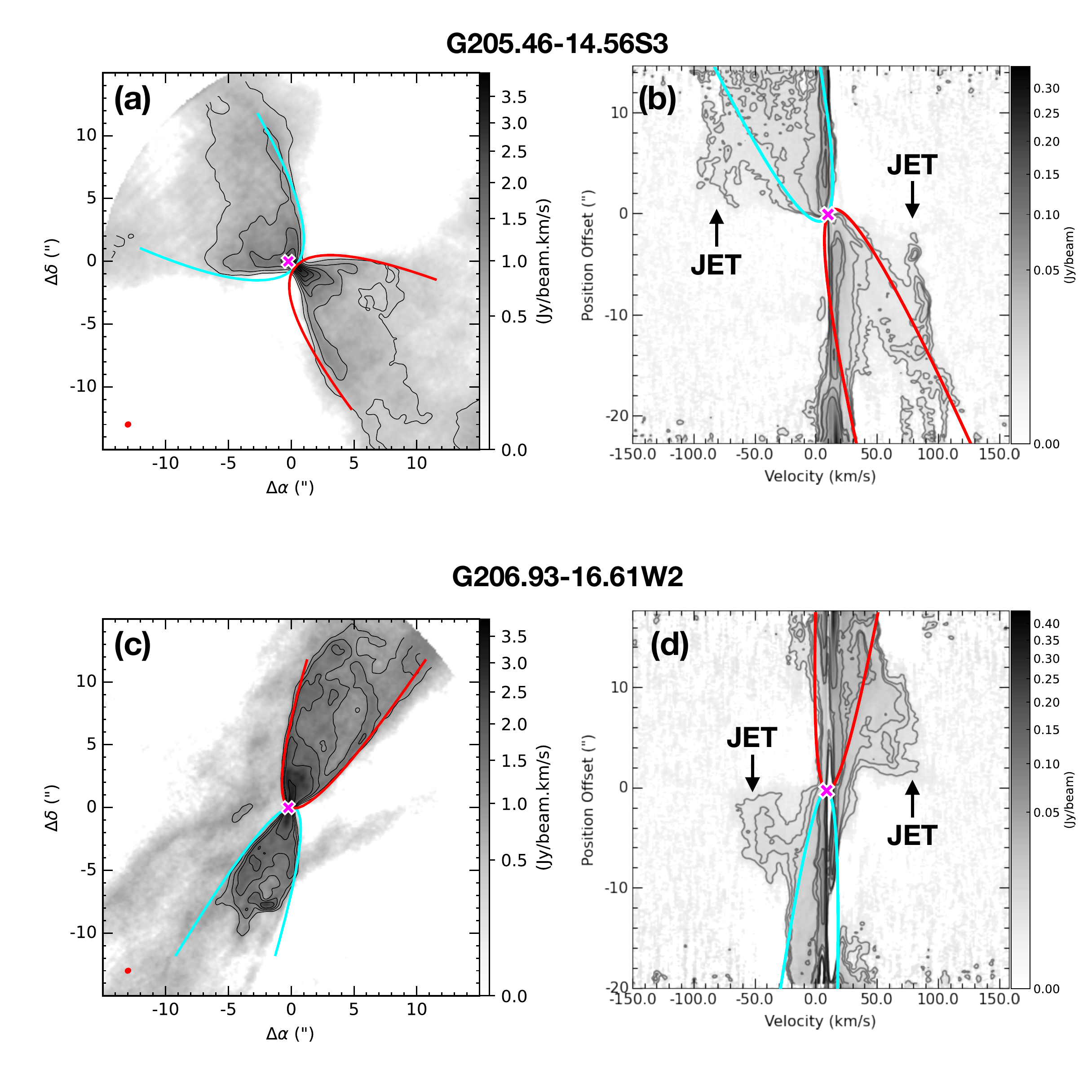}{0.90\textwidth}{}
\caption{
(a) ALMA $^{12}$CO(2$-$1) map for G205S3. The black contours are at 3n$\sigma$, where n = 1, 2, 3, ...and $\sigma$ = 0.16 Jy/beam.km/s. The best fit parabolas  are shown for blushifted (in blue) and redshifted (in red) outflow shell.  The cross mark represents the source location.
(b) Position-velocity (PV) diagram along the jet axis. The contours are at 3$\times$(1, 2, 3, 10, 16)$\sigma$, where $\sigma$ = 0.0010 Jy\,beam$^{-1}$. The parabolas in blushifted (in blue) and redshifted (in red) components are best fits of the outflow shell for the corresponding estimated C values estimated in panel `a', with the best fit $i$ = 40$\degr$ (see text for details). The cross mark at (V$_{sys}$, 0) represent the source location. The high-velocity components are marked as ``JET". 
(c) $^{12}$CO(2$-$1) map for G206W3. All the symbols have the same meaning as panel `a', and the sensitivity is $\sigma$ =  0.30 Jy/beam.km/s.
(d) PV diagram for G206W3 with the sensitivity of $\sigma$ = 0.0011 Jy\,beam$^{-1}$. All the symbols have the same meaning as panel `b'. The best fit inclination angle is $i$ = 10$\degr$ for both lobes. 
}
\label{fig:CO_PV_G205S3_G206W2}
\end{figure*} 
Figures \ref{fig:G205S3_G206W2SiO_jet}b and d show the position-velocity (PV) diagrams of the SiO emission traced along each jet-axis direction. The systemic velocities are obtained from peak C$^{18}$O emission and are marked in Figures \ref{fig:G205S3_G206W2SiO_jet}b and d (column\,3 of Table~\ref{tab:outflow_jet}). G205S3 and G206W2 exhibit line-of-sight projected maximum velocities of V$_{max,SiO}$ $\sim$ 120 and 106 km\,s$^{-1}$, respectively  (column\,5 of  Table~\ref{tab:outflow_jet}). 

The jet from G205S3 is highly collimated, whereas that from G206W2 is relatively extended spatially.  Both jets exhibit knots (marked as B1, B2, ... in the blue lobe and R1, R2.... in red lobe), which are possibly formed by a chain of internal shocks via a semi-periodic variation in the jet velocity \citep[][]{2016MNRAS.460.1829M,2019ApJ...874...31W}. 
We estimate mean line-of-sight jet velocities from the peak positions of the knot structures, yielding V$_{j,obs}$ $\sim$ 80 and 50 km\,s$^{-1}$ for G205S3 and G206W2, respectively. Considering projection effects due to the inclination angle  ($i$), the corrected velocities (V$_{j, corr}$) are related as  V$_{j,obs}$ = V$_{j, corr}$ $\sin(i)$. We evaluate the mean plane-of-sky separations for consecutive knots of  $\Delta$r$_{obs}$ = 2$\farcs$7 and 2$\farcs$5 for G205S3  and  G206W2, respectively. These measurements can expressed as the corrected distances ($\Delta$r$_{corr}$) using the  projection angles,  $\Delta$r$_{obs}$ = $\Delta$r$_{corr}$ $\cos(i)$.

\begin{deluxetable*}{c@{\extracolsep{1pt}}c@{\extracolsep{1pt}}c@{\hskip 1.0cm}c@{\extracolsep{1pt}}c@{\extracolsep{1pt}}cc@{\extracolsep{1pt}}c@{\extracolsep{1pt}}c@{\extracolsep{1pt}}cc}[h]
\tablecaption{Outflow and Jet parameters\label{tab:outflow_jet}}
\tablewidth{0pt}
\tablehead{
\multicolumn{1}{c}{} &\multicolumn{1}{c}{}& \multicolumn{1}{c}{C$^{18}$O}&  \multicolumn{3}{c}{SiO}&   \multicolumn{4}{c}{CO}\\
\cline{3-3}  \cline{5-6} \cline{7-11}\\ 
\multicolumn{7}{c}{}&   \multicolumn{2}{c}{} & \multicolumn{2}{c}{corrected$^\dagger$}\\
 \cline{10-11}\\
\colhead{Object} & \colhead{HOPS} & \colhead{V$_{sys}$} & \colhead{Lobe}   & \colhead{V$_{max,SiO}^*$} & \colhead{N$_{SiO}$} & \colhead{V$_{max,CO}$} &\colhead{$i$}& \colhead{N$_{CO}$}  & \colhead{(\.{M})} & \colhead{F$_{CO}$}\\
\colhead{} & \colhead{Id} &\colhead{(km s$^{-1}$)}   &\colhead{} & \colhead{(km s$^{-1}$)} & \colhead{(10$^{15}$ cm$^{-2}$)} & \colhead{(km s$^{-1}$)} & \colhead{(deg)}  &\colhead{(10$^{17}$ cm$^{-2}$)}  & \colhead{(10$^{-6}$M$_\sun$/yr)} & \colhead{(10$^{-5}$M$_\sun$km s$^{-1}$/yr)} \\
}
\startdata
G205S3 & 315 & 9.9 & Blue& 106.7 & 0.37 & 109.5 & 40$\pm$8 & 0.95  & 0.29 & 2.16\\
(Class I)  &  &   & Red & 120.1 & 0.22 & 96.3 & 40$\pm$8 & 0.96 & 0.30 & 1.10\\
\cline{1-11}      
G206W2 & 399 & 8.8 & Blue& 81.8 & 1.30 & 76.4   & 10$\pm$5  & 1.50 & 2.88 & 9.5 \\
(Class 0) & &    & Red & 103.0 & 1.21 & 76.2  & 10$\pm$5  & 1.34 & 2.66 & 28.84 \\       
\enddata
\tablecomments{$^*$ The mean observed jet-velocities for G205S3 \& G206W2 are estimated as V$_{j,obs}$\,$\sim$\,80 and 50 km\,s$^{-1}$, respectively.\\ 
$^\dagger$ Corrected for inclination angles ($i$). 
}
\end{deluxetable*}

\subsection{Outflow shell in CO: Inclination angle}\label{sec:results_CO_inclination}
Following the simple analytical model by \citet[][]{2000ApJ...542..925L}, we determine the  physical structure of the outflow using the CO emission.  In this model, the molecular outflow can be represented as a radially expanding parabolic shell driven by the underlying wide-angle wind. In a cylindrical coordinate system, the outflow shell can be described by the equation $z = CR^2$, where $z$ is the outflow axis and $R$ is the radial distance from that axis \citep[for a schematic diagram, see Figure 21 by][]{2000ApJ...542..925L}.

Figure \ref{fig:CO_PV_G205S3_G206W2}a displays the CO outflow components of G205S3. We fitted a parabola to the outermost contour closer to the real outflow cavity wall, yielding C = 0.20 and 0.26 for the blue and red lobe, respectively. Using these `C' values, we then fitted parabolas for the low-velocity outflow cavities in the PV diagram (Figure \ref{fig:CO_PV_G205S3_G206W2}b) for  the corresponding outflow lobes, which provided  best fits of inclination angle, $i$ = 40$\degr$ $\pm$ 8$\degr$ for both the lobes. Notice that the apparent continuum center for this object is shifted in position from the center of the `neck' towards the blue lobe due to the projection effect (Figure \ref{fig:CO_PV_G205S3_G206W2}a).  
Similarly, we estimated `C' values of 0.90 and 0.80  (Figure \ref{fig:CO_PV_G205S3_G206W2}c) for the blue and red lobes of G206W2, respectively. The best fitted parabolas in the PV diagram for G206W2 provide an inclination of  $i$ = 10$\degr$ $\pm$ 5$\degr$ for both lobes (Figure \ref{fig:CO_PV_G205S3_G206W2}d). 
Here, we note that our fits ignore the likely high-velocity jet-structure (marked as ``JET" in Figures \ref{fig:CO_PV_G205S3_G206W2}b, and d) in the PV diagram, since the model of \citep[][]{2000ApJ...542..925L} mainly describes the low-velocity winds. We did not incorporate with the jet-driven shell model, since no convex spur-like structures were observed in the PV diagram driven by the pulsating jet, as seen in the case of HH\,212 \citep[see Figure 13 by][]{2000ApJ...542..925L}.

\subsection{Jet mass-loss rate}\label{sec:results_jetmass}
Assuming optically thin CO emission, beam-averaged CO column densities (N$_{CO}$) were estimated from the high-velocity channels of each lobe of each source. Here, the high-velocity channels are defined as $|V-V_{sys}| >$ 65 km\,s$^{-1}$ for G205S3 and $>$ 45 km\,s$^{-1}$ for G206W2 (Figures \ref{fig:G205S3_G206W2SiO_jet}a and c) up to a maximum velocity CO emission (V$_{max,CO}$ in Table \ref{tab:outflow_jet}). The lower limits of the high-velocities are adopted from the SiO emission which include most of the jet emission. The mean intensity, integrated over velocity, is then obtained from the whole area encompassed by the jet emission. Assuming a specific excitation temperature of 150 K within the jet, the integrated intensity is converted into N$_{CO}$.
The beam-averaged H$_2$ column density N$_{H_2}$ is then derived assuming X$_{CO}$\footnote{It might be as small as 10$^{-4}$. In that case, the mass-loss rates would be increased by a factor of 4 \citep[][]{2010ApJ...717...58H}. The above estimation can thus be considered as the lower limit for the mass-loss rate. 
} = N$_{CO}$/N$_{H_2}$ = 4 $\times$ 10$^{-4}$ \citep[][]{1991ApJ...373..254G}. The jet mass-loss rate (\.{M}$_j$) is obtained by assuming the jet as a uniform cylinder of gas flowing at constant density and speed along the jet axis over the transverse beam direction. 
Thus, \.{M}$_j$ within a single jet lobe is determined from\\\\ 
\begin{equation}\label{equ:mass_equation}
\indent
\dot{M_j} = \mu_{H_2} m_H \frac{N_{CO}}{X_{CO}} V_{j,obs} b_m,\\\\
\end{equation}
where the mean molecular weight $\mu_{H2}$ = 2.8 
\citep[e.g.][]{2008A&A...487..993K}, $m_H$ is the mass of the hydrogen atom, and V$_{j,obs}$ is the observed mean jet velocity (section \ref{sec:results_SiO}).  
  The beam size ($b_m$) transverse to the jet direction is assumed to be the jet width, since the jet is not resolved in the present spatial resolution.

The estimated values of N$_{CO}$ and \.{M}$_j$ are listed in Table \ref{tab:outflow_jet}, without and with the inclination correction by using V$_{j,obs}$ and  V$_{j,corr}$, respectively. With these calculations, the total jet mass-loss rates are 0.59 $\times$ 10$^{-6}$  and 5.54 $\times$ 10$^{-6}$ M$_\sun$ yr$^{-1}$ for G205S3 and G206W2, respectively. Such mass-loss rates are quite comparable to other protostars observed in CALYPSO survey \citep[][]{2021A&A...648A..45P}.

\begin{figure*}
\fig{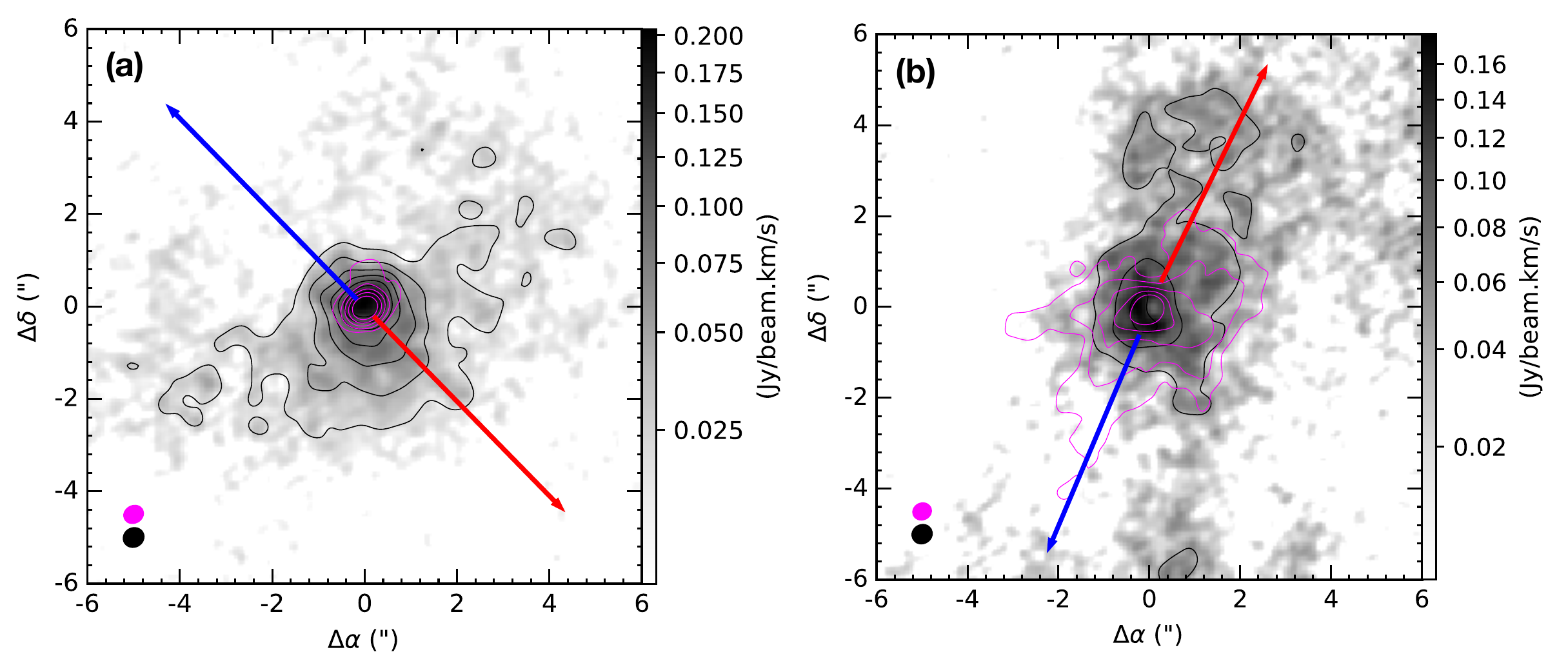}{0.95\textwidth}{}
\caption{ALMA C$^{18}$O maps for G205S3 (a) and G206W2 (b). The black C$^{18}$O contours are at  3n$\sigma$, where n = 2, 3, 4, ...and $\sigma$ = 0.007 and 0.015 Jy/beam.km/s in panel (a) and (b) respectively. The magenta contours are the 1.3 mm continuum at $\sum_{{n= 1, 2, 3, ..}}$3$\times$2$^n$ $\sigma$, where $\sigma$ = 8 $\times$ 10$^{-5}$ Jy\,beam$^{-1}$ (panel a) and 15 $\times$ 10$^{-5}$ Jy\,beam$^{-1}$ (panel b). The synthesized beams are shown in black (C$^{18}$O) and magenta (continuum) in the lower left. The red and blue arrows indicate the redshifted and blueshifted jet axis.}
\label{fig:C18O_G205S2_G206W2}
\end{figure*} 

\subsection{Outflow force}\label{sec:result_forcemomentum}
Outflow momentum flux or force (F$_{CO}$) is a vital ingredient to evaluate the outflow energetics of the protostars. From our limited field-of-view, we first estimated the outflow mass for each channel from the CO emission above 3$\sigma$ using the following equation \citep[][]{2015A&A...576A.109Y}:
\begin{equation}\label{equ:outflowmass_equation}
\indent
M_{k} = \mu_{H_2} m_H A \frac{\sum_{l}{N_{CO,l}}}{X_{CO}}\\\\
\end{equation}
Here, A is the surface area of one pixel and the sum is over all outflow pixels $l$ on the k-th channel. We assume a lower mean specific excitation temperature of T$_{ex}$ = 50 K and a less CO abundance ratio of X$_{CO}$ $\sim$ 10$^{-4}$ for the outflow than that of the jet. The momentum ($P$) is calculated by multiplying each channel by the central velocity (V$_k$) of the associated channel. The momentum is then integrated over all the channels from  starting velocity (V$_{CO,min}$) to maximum velocity (V$_{CO,max}$) of CO emission over all pixels above 3$\sigma$ emission. Therefore, the F$_{CO}$ can be expressed as 
\citep[][]{2015A&A...576A.109Y}:
\begin{equation}\label{equ:force_equation}
F_{CO} = f_{ia}\frac{V_{CO,max}\sum_k{M_k V_k}}{R_{CO}}
%
\end{equation}
%
where $f_{ia}$ is the inclination correction factor and $R_{CO}$ is the length of the red and blueshifted outflow lobes. The measured values of F$_{CO}$ for different lobes are listed in Table \ref{tab:outflow_jet}. The difference in red and blue lobe outflow occurs due to low signal-to-noise ratio, surrounding environment, and  smaller field-of-view, which could not trace the whole outflow lobe in either direction and results in the apparently unequal extent of blue and red-shifted lobes. Here, we note that changing T$_{ex}$ by $\pm$ 25 K will change the estimated column densities by 10–20\%. 


\section{Discussion}\label{sec:discussion}
\subsection{SiO emission in Jet}\label{sec:disc:Sio_emission}
 
At early protostellar phases, the outflow usually exhibits a higher mass-loss rate and produces higher density material than for later phases. Hence, SiO should be a better tracer of jets from earlier phase protostars, versus later phases \citep[][]{1991ApJ...373..254G,2006ApJ...649..845S,2007A&A...468L..29C,2020ApJ...905..116S}. Therefore, the detection of SiO emission in jets is most likely indicating a younger phase of the protostars and a higher mass-loss rate. Such SiO emission is observed in several Class\,0 protostars e.g., B\,335 \citep[][]{2019ApJ...873L..21I,2019A&A...631A..64B}, 
HH\,212 \citep[][]{2017NatAs...1E.152L,2017SciA....3E2935L}, 
L1157 \citep[][]{2015A&A...573L...2T,2016A&A...593L...4P}, 
HH\,211 \citep[][]{2016ApJ...816...32J,2018ApJ...863...94L}, 
and IRAS\,04166+2706 \citep[][]{2009A&A...495..169S,2017A&A...597A.119T}.

 We observe  SiO emission in the jet of a Class\,I source, G205S3.
We estimated SiO column densities (N$_{SiO}$) for G205S3 and G206W2 following the method described in section \ref{sec:results_jetmass}, and the values are tabulated in Table \ref{tab:outflow_jet}. The SiO abundances (X$_{SiO}$ = X$_{CO}$ $\times$ N$_{SiO}$/N$_{CO}$) turn out to be 1.25 $\times$ 10$^{-6}$ and 3.5  $\times$ 10$^{-6}$ for G205S3 and G206W2, respectively. Similar types of SiO abundances were also observed in other protostellar jets in the literature \citep[e.g.][]{2021A&A...648A..45P}. The denser jet of Class\,0 source G206W2 is possibly indicating a higher accretion phase than G205S3.  

If we compare the integrated brightness temperatures of the SiO maps (Figure \ref{fig:G205S3_G206W2SiO_jet}a and c), G205S3 exhibits a fainter jet ($\sim$ 200 K.km\,s$^{-1}$) than G206W2 ($\sim$ 1000 K.km\,s$^{-1}$). For context, the Class\,0 system, HH\,212 exhibits an SiO integrated brightness temperature similar to that of G206W2 ($\sim$ 1000 K.km\,s$^{-1}$) \citep{2017NatAs...1E.152L} and has a high mass-loss rate of $\sim$ 1.1 $\times$ 10$^{-6}$ M$_\sun$ yr$^{-1}$ \citep[][]{2015ApJ...805..186L}. The fainter SiO jet indicates that G205S3 is arguably at the tail end of its  ``high-accretion" phase.

Previous high-resolution ALMA observations  ($\sim$ 8 au) of SiO jet near the base of HH\,212 unveiled that the protostellar jets removed the residual angular momenta from the innermost part ($\sim$ 0.05 AU) region \citep[][]{2017NatAs...1E.152L}. The bolometric luminosity (L$_{bol}$ $\sim$ 9 L$_\sun$) of HH\,212 suggests that SiO-jet is possibly originating from the dust sublimated zone. From our SiO and CO emission, we estimated SiO-to-CO abundances (X[SiO/CO]) are $\sim$ 4$\times$10$^{-3}$ and $\sim$ 9$\times$10$^{-3}$ for G205S3 and G206W2, respectively, which suggest that the jet is possibly launched from the dust poor zones within the dust sublimation radius.  Moreover, as Class\,0 sources evolved to Class\,I, the X[SiO/CO]) also declined with decreasing mass-loss rate \citep[][]{2020A&A...636A..60T}. Lower SiO abundance in G205S3 than G208W2 could be related to evolutionary phases, although, a statistically more number of observations of Class\,0 and Class\,I samples are required to constrain a systematic change in  X[SiO/CO] with protostellar evolution. Here we note that the SiO line could be optically thick even in the high-velocity jet \citep[e.g.,][]{2021A&A...648A..45P}. Since CO is likely optically thin in the jet, the SiO-to-CO abundance ratio derived here could constitute a lower limit to the true SiO-to-CO abundance ratio.

More specifically, comparing the mass-loss rate and X[SiO/CO]) with the astrochemical models of \citet[][]{2020A&A...636A..60T} (see their Figure 12), the SiO-rich jet of G206W2 is expected to be launched from the dust sublimation zones. The jet of G205S3 could have been launched even from the dust poor zones (dust mass fraction $\sim$ 10$^{-3}$), which could be the outer boundary of dust sublimation zones.
The bolometric luminosities of both G205S3 and G206W2 suggest that their dust sublimation radii are less than 0.1 au \citep[][]{2007prpl.conf..539M}.  Silicon is possibly released into the gas phase near the base of the jet and can synthesize SiO molecules. \citep[][]{1991ApJ...373..254G}.  For low-luminosity objects, Si$^+$ recombination and SiO formation occur faster than the photodissociation by the relatively weak ultraviolet radiation from the central protostar, yielding abundant SiO.

To summarize, we surprisingly detected SiO emission in the jet of evolved Class\,I source G205S3, as usually observed in Class\,0 sources (e.g., G206W2, HH\,212) in this study as well as in the literature. To validate the true evolutionary phase of G205S3, we investigate the aspects of possible misclassification of G205S3.

\subsection{Evolutionary status}

\subsubsection{Inclination Effect}\label{sec:disc:inclination}
The spectral classifications through T$_{bol}$ and $\alpha_{IR}$ of protostars are based on the observed fluxes of the component central source, disk, and envelope, obtained by telescopes with different spatial resolutions. 
The system's inclination angle to the line-of-sight, however, constitutes a major uncertainty to the interpretation of the observed fluxes, given the asymmetries inherent in the circumstellar environment \citep[][]{2009ApJS..181..321E,2016ApJS..224....5F}. 
For example, the face-on view of a flattened envelope-disk (or pole-on outflow) yields relatively high infrared fluxes due to less extinction along the line of sight compared with an edge-on view (pole-off outflow) of the same system. Thus, as a result, a Class\,0 source may appear to be Class\,I. Disentangling observed fluxes and inclination angles can be difficult. Fortunately, however, \citet[][]{2016ApJS..224....5F} modelled the spectral energy distributions (SEDs) of protostars at different inclination angles. Indeed, their models suggest an inclination angle of 50$\degr$ for both  sources. Although the SED-derived inclination angle of G205S3 is close  to that measured from the outflow morphology ($\sim$ 40$\degr$; Table \ref{tab:outflow_jet}), that of G206W2 is more dissimilar to the inclination angle  suggested by its outflow ($\sim$ 10$\degr$).

We applied the inclination effect when determining mass-loss rates in Table \ref{tab:outflow_jet}. Nevertheless, the mass-loss rate of G205S3 remains at the lower end of what is expected for Class\,0 systems. At present, it is not clear how much the inclination affects the spectral classification of G205S3. It is also difficult to predict how significantly the inclination of  40-60$\degr$ might expose the central source to the observer, given that the object is still not close to face-on. 
Therefore, we probe other evidence to validate the evolutionary status of the sources, such as (i) Outflow energetics, (ii) C$^{18}$O emission and surrounding material, (iii) luminosity vs mm-flux, in the next sections.

\subsubsection{Outflow Energetics}
Previous observations revealed that a Class\,I source usually exhibits smaller F$_{CO}$ than Class\,0 with similar L$_{bol}$ \citep[][]{1996A&A...311..858B,2007A&A...472..187H,2010MNRAS.408.1516C,2015A&A...576A.109Y}, which suggest that the Class\,0 sources are more energetic than the Class\,I counterpart. Assuming a CO abundance ratio of 1.2 $\times$ 10$^{-4}$, \citet[][]{2015A&A...576A.109Y} obtained the typical values of F$_{CO}$ $\sim$ 2.8 $\times$ 10$^{-5}$ M$_\sun$km s$^{-1}$/yr for Class\,I sources, and $\sim$ 6.9 $\times$ 10$^{-4}$ M$_\sun$km s$^{-1}$/yr for Class\,0 sources. We estimated a total of F$_{CO}$ $\sim$ 3.26 $\times$ 10$^{-5}$ M$_\sun$km s$^{-1}$/yr for G205S3 and $\sim$ 3.834 $\times$ 10$^{-4}$ M$_\sun$km s$^{-1}$/yr for G206W2 (Table \ref{tab:outflow_jet}). The smaller F$_{CO}$ of G205S3 make its more similar to a Class\,I source whereas the relatively high value F$_{CO}$ indicate that G206W2 is probably a Class\,0 source.

\subsubsection{C$^{18}$O emission and surrounding material}
\begin{figure*}
\fig{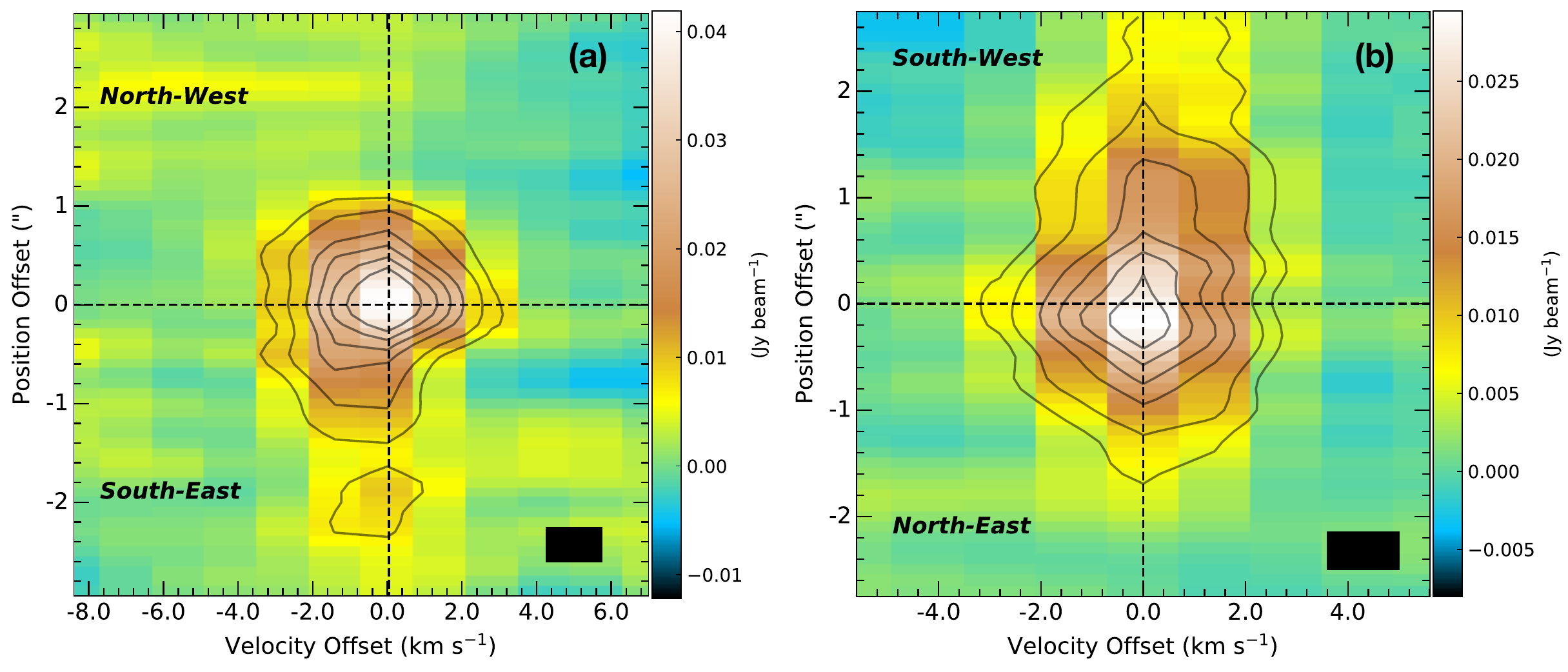}{0.95\textwidth}{}
\caption{C$^{18}$O position velocity diagram along perpendicular to jet axis for (a) G205S3 and (b) G206W2. The source location is at the center of the cross mark. The beam size and velocity resolution are shown with black rectangle at lower right of each panel.
}
\label{fig:C18OPV}
\end{figure*} 

The bolometric temperature ($\sim$ 180 K) and near-infrared spectral index ($\alpha_{IR}$ $\sim$ 0.417) suggest that G205S3 is a Class\,I source \citep[][]{2016ApJS..224....5F,2020ApJS..251...20D}.
In addition, if we compare Figure \ref{fig:C18O_G205S2_G206W2}a \& b, G205S3 displays relatively  compact C$^{18}$O and 1.33\,mm continuum emission compared with G206W2. The disk-envelope mass of G205S3, $\sim$ 0.167$\pm$0.072 M$_\sun$, is a few times smaller than G206W2, $\sim$  0.771$\pm$0.333 M$_\sun$, at the same beam resolution, $\sim$ 140 au \citep{2020ApJS..251...20D}. Such compact emission in G205S3 is consistent with a smaller envelope surrounding the protostar, with the emission most likely emanating from an evolved disk, consistent with a Class\,I object. For G206W2, the envelope dominates the observed emission, consistent with a very early, Class\,0, stage of protostellar evolution.  It is to be noted that the compactness of a source based on continuum or line images (e.g., C$^{18}$O) largely depends on the sensitivity of the maps.

To extract the size of compact components, we explore the continuum visibility amplitudes (Jy) as a function of $uv$ distance (k$\lambda$). Following \citet{2008ApJ...685.1026L}, the $uv$ visibilities were fitted with two Gaussian components. The compact component of G205S3 has a deconvolved size of $\sim$ 0\farcs20 and flux $\sim$ 30 mJy. The extended component has a deconvolved size of $\sim$ 2\farcs28 and flux $\sim$ 47 mJy. Similarly, G206W2 has a compact component of size $\sim$ 0\farcs60 (flux $\sim$ 100 mJy) and extended component $\sim$ 2\farcs55 (flux $\sim$ 320 mJy). These sizes of compact component are consistent with 1.3\,mm continuum, as estimated by \citet[][]{2020ApJS..251...20D}. From the $uv$ domain, These results also suggest that G205S3 possesses a relatively compact core, surrounded by a relatively smaller envelope.

Larger scale JCMT observations (beam size $\sim$ 5600 au) also reveal a relatively compact  core surrounding the G205S3 (size $\sim$ 0.04 pc; mass $\sim$ 0.86$\pm$0.16 M$_\sun$) compared with G206W2 (size $\sim$ 0.06 pc; mass $\sim$ 3.38$\pm$0.15 M$_\sun$) \citep[][]{2018ApJS..236...51Y}. Here, we remark that the significant amount of gas in the  core should be resolved out by the interferometric ALMA observations. Thus, the  interpretation of the compact emission distribution may depend somewhat on the spatial resolution of the envelope observations.

The C$^{18}$O PV diagram is shown in Figure \ref{fig:C18OPV}. Similar to Figure \ref{fig:C18O_G205S2_G206W2}, G205S3 appears more compact than G206W2 in position-velocity space. The C$^{18}$O emission is most likely originating from the envelope part  of both the sources. No Keplerian rotation profiles are observed clearly on both sources. Higher spatial and velocity resolution data  could efficiently determine the Keplerian rotation from the inner part and possibly allow  to measure protostellar masses \citep[e.g.,][]{2014ApJ...786..114L}, hence could constrain the evolutionary phases of the stellar cores.

\subsubsection{Luminosity and Flux correlation}
\begin{figure*}
\fig{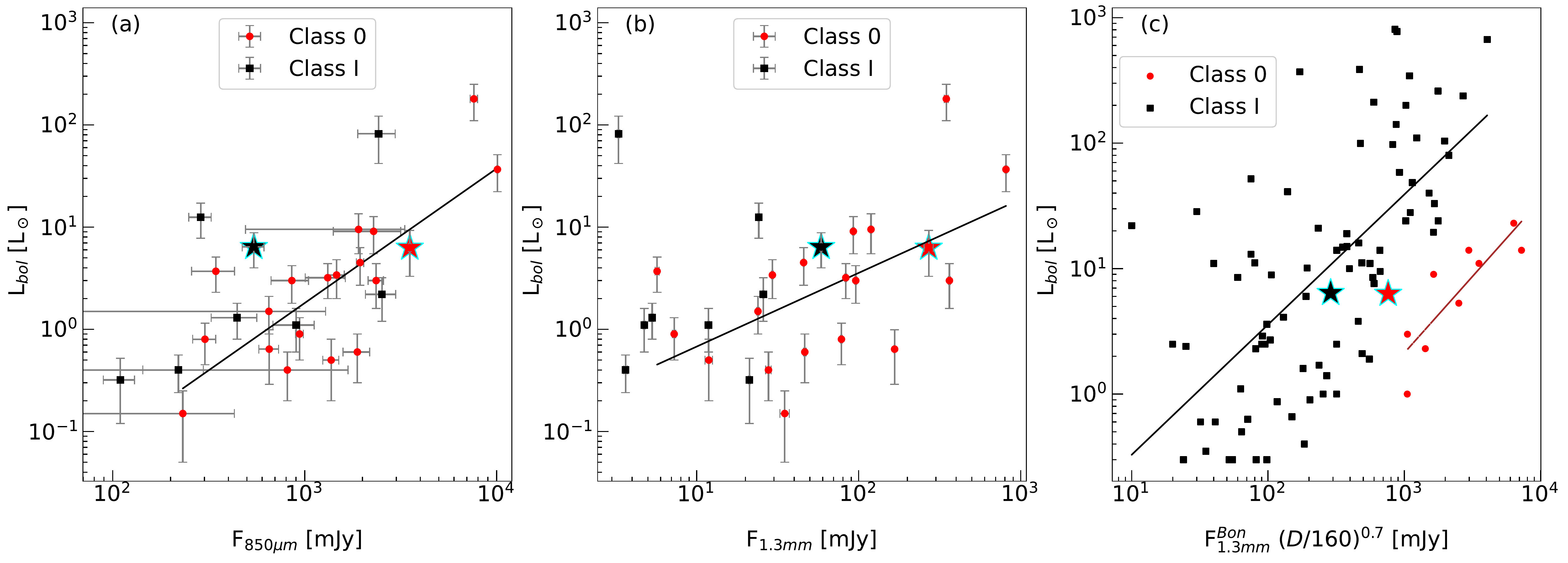}{0.95\textwidth}{}
\caption{Flux vs bolometric luminosity for Class\,0 (red circle) and Class\,I (black square), for the ALMASOP sources (panel-a and b) from \citet[][]{2020ApJS..251...20D}. The solid lines in panel-a and b indicate the best fit correlation for the Class\,0 sources only. The star marks indicate Class\,0 (G206W2 in red) and Class\,I (G205S3 in black) sources, studied in this paper. Panel-a utilizes JCMT 850 $\mu$m fluxes and panel-b utilizes the ALMA 1.3\,mm fluxes. In panel-c, G205S3 and G206W2 are compared with the Class\,0 and Class\,I sample of  \citep[][]{1996A&A...309..827S}. The solid black line is the best fit for the Class\,I and brown line is the best fir for Class\,0 sources.
}
\label{fig:Lbol_flux_Class0-1}
\end{figure*}

The (sub)millimeter and far-IR fluxes contribute more to L$_{bol}$ for  the Class\,0 than that of Class\,I sources since these fluxes are mostly originating from the envelope and Class\,0 should have larger surrounding material. In Figure \ref{fig:Lbol_flux_Class0-1}, luminosities of the Class\,0 and Class\,I sources are shown as a function of JCMT 870 $\mu$m (panel a) and ALMA 1.3\,mm (panel b) fluxes for the non-multiple\footnote{Luminosity of multiple systems is contributed from the different protostellar components within the common envelope. To reduce the uncertainty in the correlation, we removed the known multiple sources from this analysis.} ALMASOP sources \citep[see][]{2020ApJS..251...20D}. A linear fit only for the Class\,0 sources are also shown in both panels. No obvious correlation was found between the Class\,0 and Class\,I sources.  One interesting fact is that most of the Class\,I sources, $\sim$ 75\% for 850$\mu$m (panel a) and $\sim$ 90\%  for 1.3\,mm (panel b), are located above the locus of Class\,0 sources (black fitted line). Our studied targets, G205S3 (Class\,I: L$_{bol}$ = 6.4$\pm$2.4 L$_\sun$) and G206W2 (Class\,0: L$_{bol}$ = 6.3$\pm$3.0 L$_\sun$),  have similar type of luminosities. However, if we see their location in Figure \ref{fig:Lbol_flux_Class0-1}, G205S3 (black star mark) exhibit lower fluxes than G206W2 (red star mark) in both 870 $\mu$m (panel a) and 1.3\,mm emission (panel b). Here, we note that the use of single-dish data (JCMT) would reduce the contamination by a possible massive disk in Class\,I sources than interferometer  (ALMA), and the low-resolution JCMT data could include significant emission from the larger scale cloud emission beyond the envelope and presence of possible multiple sources. Both could potentially affect the interpretation of flux-luminosity correlation.

A similar type of correlations were also studied for 1.3\,mm fluxes of Class\,0 and Class\,I sources in \citet[][]{1993A&A...273..221R} and \citet[][]{1996A&A...309..827S}. We adopted the single-dish observations of Class\,0 and Class\,I sample from \citet[][]{1996A&A...309..827S} in Figure \ref{fig:Lbol_flux_Class0-1}c. We scale down 870$\mu$m JCMT single-dish observations of G205S3 (black star mark) and G206W2 (red star mark) from \citet[][]{2020ApJS..251...20D} to 1.3\,mm fluxes with spectral index 3.0 and 160 pc distance with a distance correlation from \citet[][]{1996A&A...309..827S}. With their limited number of Class\,0 sample, G205S3 is quite convincingly located in Class\,I cluster and G206W2 is located near the  Class\,0 cluster. 
Such correlations are quite obvious and could be utilized to distinguish between different spectral classes. More robust single-dish observations at 1.3\,mm of a larger sample of Class\,I and Class\,0 sources are needed to constrain the correlation.

\subsection{Jets Knots and Accretion Variability }\label{sec:disc:var}

Variable accretion may be due to a circumbinary disk around a close binary, 
potentially affecting the final mass-budget of the binary and planet formation \citep[][]{2002A&A...387..550G,2012ApJ...749..118S}. 
Periodic variations due to pulsed accretion have been observed in T-Tauri binary systems (\citealp[e.g.\ DQ\,Tau:][]{1997AJ....113.1841M}; \citealp[UZ\,Tau\,E:][]{2007AJ....134..241J}). 
Another Class\,I, possibly binary, system LRLL 54361 has pulsed accretion with a high-accretion rate of 3$\times$10$^{-6}$ 
M$_\sun$/yr \citep[][]{2013Natur.493..378M}.

The JCMT Transient Survey \citep{herczeg17} has been monitoring Class\,0 and I protostars in submillimeter continuum for over four years, with monthly cadence \citep[]{johnstone18,2021arXiv210710750L}. 
G205S3 is a JCMT-monitored variable, with an extrapolated episode timescale $\sim 8\,$yrs, and an implied 10\% mass accretion rate variability \citep[][]{2021arXiv210710750L}. Similar variability properties are found in the mid-Infrared (mid-IR) \citep[][]{2021arXiv210710751P}. G206W2 is not included in the JCMT sample and it is too bright for inclusion in the mid-IR analysis by \citet[][]{2021arXiv210710751P}. Extracting the light curve from the NEOWISE catalogue (Wooseok Park, private communication) reveals a long-timescale secular change in brightness.

The G205S3 SiO jet knots imply ejection episodes 54$^{+30}_{-10}$ years apart ($\Delta$r$_{corr}$/V$_{j, corr}$; section \ref{sec:results_SiO}), longer than the extrapolated accretion event uncovered by the JCMT. A statistical ensemble analysis of the JCMT protostars suggests that these burst events typically occur every $\sim$50 years, consistent with the G205S3 knot separation.  
Repetitive ejection may be related to episodic accretion of dense clumps created by instabilities within a massive disk or magnetorotational instabilities \citep{machida11, bae14,vorobyov15}. Alternatively, embedded planetesimals or a circumbinary disk due to an unresolved binary  might create periodic accretion signatures \citep{bonnell92,munoz16}. However, observationally no binary has been detected in the G205S3 system so far. 
G206W2 exhibits ejections with a 17$^{+30}_{-10}$ year timescale, more closely matching the extrapolated mid-IR light curve. Further high spatial resolution outflow studies and continued mid-IR through submillimeter monitoring are required to better diagnose the underlying accretion-ejection scenarios.

\section{Conclusions}
We detect SiO jet emission and a moderately high mass-loss rate of $\sim$ 0.59 $\times$ 10$^{-6}$ M$_\sun$ yr$^{-1}$ for the isolated Class\,I source, G205S3.  These outflow features are more typically seen in younger, Class\,0 sources like the representative source G206W2 (\.{M}$_j$ $\sim$ 5.54 $\times$ 10$^{-6}$ M$_\sun$ yr$^{-1}$). We estimated a high inclination ($i$ $\sim$ 40$\degr$) of G205S3. Perhaps G205S3 has been misidentified as Class\,I from the SED of observed fluxes due to its high inclination. However, 
the compact 1.3\,mm continuum, C$^{18}$O emission, location in the bolometric luminosity and sub-millimeter fluxes diagram, lower SiO-to-CO abundances ($\sim$ 4$\times$10$^{-3}$) in the jet are consistent with a Class\,I system.  The SiO-to-CO abundances, mass-loss rate, and bolometric luminosity indicate that the jets from both the sources, G205S3 and G206W2, are possibly launched from the innermost dust poor zones. The measured values of outflow forces, F$_{CO}$ $\sim$ 3.26 $\times$ 10$^{-5}$ M$_\sun$km s$^{-1}$/yr for G205S3 and $\sim$ 3.834 $\times$ 10$^{-4}$ M$_\sun$km s$^{-1}$/yr for G206W2 infer that G205S3 could be a Class\,I  whereas G206W2 is possibly a Class\,0 source. The episodic ejection could be due to the presence of an unknown binary, a planetary companion, or dense clumps in a circumbinary disk, which also could trigger such high accretion. Alternatively, some special environmental conditions or rare orbital parameters may be in play that favors the high accretion rate. Further high-resolution Near-Infrared and ALMA observations are needed.

\acknowledgments
We thank the anonymous referee for the constructive comments.  This paper makes use of the following ALMA data:  ADS/JAO.ALMA$\#$2018.1.00302.S. ALMA is a partnership of ESO (representing its member states), NSF (USA) and NINS (Japan), together with NRC (Canada), NSC and ASIAA (Taiwan), and KASI (Republic of Korea), in cooperation with the Republic of Chile. The Joint ALMA Observatory is operated by ESO, AUI/NRAO and NAOJ.  S.D. and C.-F.L. acknowledge grants from the Ministry of Science and Technology of Taiwan (MoST 107-2119-M- 001- 040-MY3) and the Academia Sinica (Investigator Award AS-IA-108-M01). C.W.L. is supported by the Basic Science Research Program through the National Research Foundation of Korea (NRF) funded by the Ministry of Education, Science and Technology (NRF-2019R1A2C1010851).  D.J. is supported by the National Research Council of Canada and by a Natural Sciences and Engineering Research Council of Canada (NSERC) Discovery Grant.  Tie Liu is supported by international partnership program of Chinese academy of sciences grant No.114231KYSB20200009 and the initial fund of scientific research for high-level talents at Shanghai Astronomical Observatory. P.S. was partially supported by a Grant-in-Aid for Scientific Research (KAKENHI Number 18H01259) of the Japan Society for the Promotion of Science (JSPS). L.B. acknowledges support from ANID project Basal AFB-170002.\\\\

\facility{ALMA}\\

\software{Astropy \citep[][]{2013A&A...558A..33A}, APLpy \citep[][]{2012ascl.soft08017R}, Matplotlib \citep[][]{Hunter:2007}, CASA \citep[][]{2007ASPC..376..127M}}


    
\bibliography{sample63}{}
\bibliographystyle{aasjournal}

\end{document}